\title{Optimality of Non-Restarting CUSUM charts}
\author{F. Din-Houn Lau \qquad Axel Gandy
  \\Department of Mathematics, Imperial College London}
\date{}

\documentclass[11pt]{article}
\usepackage{amsmath,amssymb}
\usepackage{amsfonts}
\usepackage{graphicx}
\usepackage[margin=1.1in]{geometry}
\usepackage{bm}
\usepackage{natbib}
 \bibpunct{(}{)}{;}{a}{,}{,}
\usepackage{subfig}

\DeclareMathOperator*{\esssup}{ess sup}
\newcommand{\mc}[1]{\mathcal{#1}}
\newcommand{\mb}[1]{\mathbb{#1}}

\newcommand{\mbf}[1]{\mathbf{#1}}
\newtheorem{thm}{Theorem}
\newtheorem{define}{Definition}

\begin{document}
\maketitle

\begin{abstract}
  We show optimality, in a well-defined sense, using cumulative sum
  (CUSUM) charts for detecting changes in distributions. We consider a
  setting with multiple changes between two known distributions. This
  result advocates the use of non-restarting CUSUM charts with an
  upper boundary. Typically, after signalling, a CUSUM chart is
  restarted by setting it to some value below the threshold. A
  non-restarting CUSUM chart is not reset after signalling; thus is
  able to signal continuously. Imposing an upper boundary prevents the
  CUSUM chart rising too high, which facilitates detection in our
  setting. We discuss, via simulations, how the choice of the upper
  boundary changes the signals made by the non-restarting CUSUM
  charts.
\end{abstract}


\section{Introduction}
In statistical process control, one of the most commonly used control
charts is the cumulative sum (CUSUM) chart developed by
\cite{page}. As a simple example of a CUSUM chart consider the
following. Assume we sequentially observe the independent random
variables $X_t$ ($t\in\mb{N}=\{1,2,\dots\}$). Suppose the observations
$X_1,\dots,X_{n-1}$ each have distribution $N(0,1)$ and
$X_n,X_{n+1},\dots$ have $N(\Delta,1)$ for some known
$\Delta>0$. The aim is then to find the unknown change point
$n\in\mb{N}$. The classic CUSUM chart is
\begin{equation*}
  S_t=\max\{S_{t-1}+X_t-\Delta/2,0  \},\quad S_0=0.
\end{equation*}
The chart signals a change at time $\inf\{t>0;S_t\geq \alpha \}$ for
some threshold $\alpha>0$. At this time the chart suspects a change in
distribution. When the CUSUM chart crosses the threshold $\alpha$, it
is restarted in some fashion. The chart is typically reset at $0$,
with some practitioners opting for the headstart feature
\citep{Lucas}, where the chart is reset at a different value such as
$\alpha/2$. CUSUM charts were initially designed for use in industry
\citep{page}, where restarting coincides with a machine being repaired
and reset.

We are concerned with situations where the observations can switch
multiple times between two known distributions and where resetting is
not possible. For example, medical settings where ``machines'' such as
hospitals cannot be reset when a deterioration of performance is
suspected. Thus we shall use the CUSUM charts of the form
\begin{equation}\label{eq:our_CUSUM}
  R_t=f_t(R_{t-1})\quad\text{where}\quad f_t(x)=\min\{\max(x+\log\ell(X_t),0),h\}
\end{equation}
for $t\in\mb{N}$, where $\ell(x)$ is the Radon-Nikodym derivative of
$F_1$ with respect to $F_0$ and $h>0$ is a constant specifying an
upper boundary. As explained in \cite{gandy11CUSUMFDR}, a single
non-restarting CUSUM chart \eqref{eq:our_CUSUM} with $R_0=0$ is
appropriate in settings where restarting is not possible. Moreover, it
is shown that the imposition of an upper boundary facilitates the
detection when switching between an in-control state and
out-of-control state many times. The CUSUM charts defined by
\eqref{eq:our_CUSUM} constitutes a family of charts, each member
represented by a specific choice of starting value $R_0\in [0,h]$. We
distinguish two particular CUSUM charts which we shall use throughout
this paper.
\begin{define}
  For a specified upper boundary $h>0$, denote the CUSUM chart of the
  form \eqref{eq:our_CUSUM} with $R_0=0$ as $R^L_t$ and with $R_0=h$
  as $R^U_t$.
\end{define}
These are the ``extreme'' charts in \eqref{eq:our_CUSUM} i.e.\ for any
$R_0\in[0,h]$ we have $R^L_t\leq R_t \leq R^U_t$ for all $t\in\mb{N}$.

In this paper, it is shown that using CUSUM charts of the form
\eqref{eq:our_CUSUM} are optimal in a well-defined sense.  The
optimality criteria we consider is similar to that used by
\cite{lorden}. Our optimality is a generalisation of the
\cite{moustakides} result, where a single change in distribution is
considered. Showing optimality of CUSUM charts is a subject of
interest in the change point detection literature. For instance,
\cite{Poor1998} extends the optimality in the sense of \cite{lorden},
to include a exponential penalty for delay. Moreover, extensions to
continuous time processes \citep{Mous2004}, use of dependent
observations such as Markov chains \citep{Yakir1994} and random
processes \citep{Mous1998} have also been explored. However, to our
knowledge, the optimality of CUSUM charts in a setting where
observations can switch between two known distributions multiple times
has not been investigated.

To make clear our setting and the charts we shall being using,
consider the following illustrative example. Let $X_t\sim N(-1/2,1)$
when in the in-control state and $X_t\sim N(1/2,1)$ when in the
out-of-control state at time $t$. In this example, we use the charts
$R_t^L$ and $R_t^U$ with $h=16$ with threshold $\alpha=8$. These
non-restarting CUSUM charts will signal out-of-control at time $t$ if
$R_t^L\geq \alpha$. Moreover, these charts will continually signal
out-of-control whilst $R_t^L$ remains above $\alpha$. Similarly, for
signalling in-control when the upper chart $R^U_t$ drops below
$\alpha$.

\begin{figure}[tbp]
  \centering
  \includegraphics[width=\linewidth]{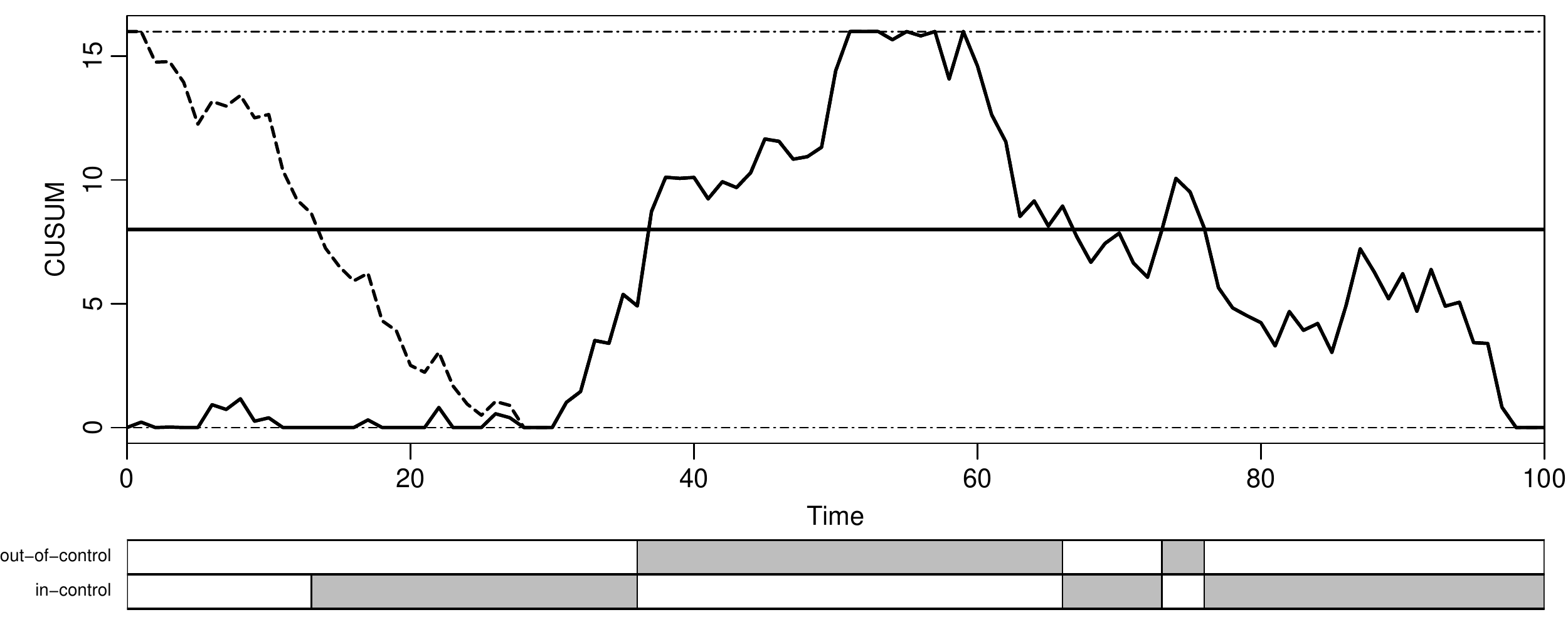}
  \caption{Two non-restarting CUSUM charts (above) with signal indicator (below).}
  \label{fig:two_eg1}
\end{figure}

In Figure \ref{fig:two_eg1} it is clear that the lower boundary at $0$
and upper boundary at $h$ are ``holding barriers'' that prevent the
chart dropping too low or rising too high. Beneath the plot of the
CUSUM chart in Figure \ref{fig:two_eg1}, is the signal indicator. The
grey areas represent the charts signalling out-of-control and
in-control, where it is clear that continuous signals are made. This
would not be the case when using a traditional CUSUM chart (i.e.\ a
chart of the form \eqref{eq:our_CUSUM} with $h=\infty$ starting at
$0$) that restarts when its threshold is crossed.

In Section \ref{sec:notation} we begin by introducing notation
appropriate in our setting. We then introduce stochastic processes as
signal processes, that allows continuous signalling. In Section
\ref{sec:mous} we present \cite{lorden} criterion of optimality as
used in \cite{moustakides}. Section \ref{sec:main} contains the main
theoretical result of the paper. Here, it is shown that using
non-restarting CUSUM charts with an upper boundary are optimal in a
well-defined sense. Although the upper boundary is introduced in the
theoretical result, its value remains unspecified. In Section
\ref{sec:prac} we explore how the choice of the upper boundary affects
the signals made and practical consequences. We conclude in Section
\ref{sec:discussion} we some remarks concerning future topics of
research.

\section{Optimality}\label{sec:optimality}
In this section we present our optimality criteria. We present the
theoretical result proving that using non-restarting CUSUM charts with
an upper boundary, are optimal in the sense defined. We begin by
introducing notation and recalling the \cite{moustakides} result.

\subsection{Notation}\label{sec:notation}
Let $(\Omega,\mc{F})$ be a measurable space with random variables
$(X_t:t\in\mb{N})$. Let $\mc{P}$ be a set of probability measures on
$(\Omega,\mc{F})$ such that for all $P\in\mc{P}$, $X_t$ ($t\in\mb{N}$)
are independent and the distribution of $X_t$,
$\mc{L}(X_t)\in\{F_0,F_1 \}$ for all $t\in\mb{N}$. Both $F_0$ and
$F_1$ are assumed to be known and mutually absolutely continuous. We
will assume that $\ell(X_1)$ has no atoms with respect to $P\in\mc{P}$
such that $\mc{L}(X_t)=F_0$ for all $t\in\mb{N}$. The $\sigma$-algebra
generated by $\{X_1,\dots,X_t \}$ is denoted by $\mc{F}_t$.
Henceforth, whenever we use an expectation $E$ or essential supremum,
we shall specify with respect to which probability measure in $\mc{P}$
by defining the distributions of the $X_t$ $(t\in\mb{N})$. We shall
refer to the case $\mc{L}(X_t)=F_0$ as $X_t$ being in the in-control
state and $\mc{L}(X_t)=F_1$ as $X_t$ being in the out-of-control state.

As we are concerned with detecting periods where the observations are
in the in-control state or the out-of-control state, it is not
convenient to work with stopping times as used in \cite{moustakides}
and \cite{lorden}. Thus we shall use a stochastic process as a signal
process. 
\begin{define}
  A signal process is a stochastic process $Z=\{Z_t : t\in\mb{N}\}$
  such that $Z_t\in\left\{0,1,\emptyset \right\}$ and $Z_t$ is
  $\mc{F}_t$-measurable for all $t\in\mb{N}$.
\end{define}
For a signal process $Z$, the events $\{Z_t =j \}$ correspond to
signalling in-control ($j=0$), signalling out-of-control ($j=1$) and
no signal ($j=\emptyset$) at time $t$. The first in-control ($j=0$) or
out-of-control ($j=1$) signal by the signal process $Z$ after time $n$
is denoted by $\tau_n^j(Z)=\inf\{t\geq n : Z_t =j \}$.

Lastly, we define CUSUM chart developed by \cite{page} $S_t^L$
and its analogue, with the roles of $F_0$ and $F_1$ swapped $S_t^U$ as
follows.
\begin{define}
  For $h=\infty$ and $f_t$ as in \eqref{eq:our_CUSUM} let, for all $t\in\mb{N}$
  \begin{equation*}
    S_t^L=f_t(S_{t-1}^L)\quad\text{and}\quad S_t^U=-f_t(-S_{t-1}^U)\quad\text{with}\quad S_0^L=S_0^U=0.
  \end{equation*}
\end{define}
\subsection{\cite{moustakides} Optimality Result}\label{sec:mous}
In this section we present the optimality result of \cite{moustakides}
in terms of signal processes. First, we recall optimality criteria
used in \cite{lorden}.
\begin{define}
  Let $P\in\mc{P}$ be such that $\mc{L}(X_t)=F_{1-j}$ for all $t< n$
  and $\mc{L}(X_t)=F_j$ for all $t\geq n$. For a signal process $Z$
  and $j\in\{0,1\}$ let
  \begin{align*}
    \mbf{D}_n^j(Z)&=\esssup E\left\{\left[\tau_1^j(Z)-n+1\right]^{+}\mid\mc{F}_{n-1}\right\},\\
    \mbf{D}^j(Z)&=\sup_{n\in\mb{N}}\mbf{D}_n^j(Z).
  \end{align*}
\end{define}
The term $\mbf{D}^j(Z)$ is the longest average delay of signalling the
$X$'s are $F_j$ distributed, guaranteed regardless of the behaviour of
the $F_{1-j}$ distributed $X$'s before the change.

For a constant $\gamma_1>0$, consider the optimization problem posed
in \cite{moustakides}
\begin{equation}\label{eq:mous}
  \begin{cases}
    \mbf{D}^1(Z)\rightarrow\min\\
    E\{\tau_1^1(Z)\mid\mc{L}(X_t)=F_1\forall t\in\mb{N}\}\geq\gamma_1
  \end{cases}
\end{equation}
where $Z$ is a signal process. As proved in \citet[][Proposition
2]{Ritov1990} a solution of \eqref{eq:mous} is the signal process
$Z^1$ where $Z^1_t=1$ if $S_t^L\geq k_L$ for $t\in\mb{N}$, where
$k_L>0$ is a constant determined by $\gamma_1$. We appeal to this
alternative proof of \cite{moustakides} to avoid complications with
the $\sigma$-algebras. By swapping the roles of $F_0$ and $F_1$ we
obtain an analogous result. More precisely, for a constant
$\gamma_0>0$, consider the optimization problem
\begin{equation}\label{eq:mous_alt}
  \begin{cases}
    \mbf{D}^0(Z)\rightarrow\min\\
    E\{\tau_1^0(Z)\mid\mc{L}(X_t)=F_0\forall t\in\mb{N}\}\geq\gamma_0
  \end{cases}
\end{equation}
where $Z$ is a signal process. A solution of \eqref{eq:mous_alt} is
the signal process $Z^0$ where $Z^0_t=0$ if $S_t^U\geq k_U$ for
$t\in\mb{N}$, where $k_U>0$ is a constant determined by $\gamma_0$.

The expectation in \eqref{eq:mous} is average time until signalling
out-of-control, when all the observations are truly in-control and
vice versa for the expectation in \eqref{eq:mous_alt}. This is
referred to as the average in-control and out-of-control run lengths
respectively.

\subsection{Main Result}\label{sec:main}
In this section we present our optimality result. We begin by defining
our optimality criteria.
\begin{define}
  For a signal process $Z$ and $j\in\{0,1 \}$ let
  \begin{align*} \nonumber 
    \mbf{C}_n^j(Z)&=\max\left\{\esssup E\left(\tau_n^j(Z)-n+1\mid \mc{F}_{n-1} \right):P\!\in\!\mc{P},\mc{L}(X_t)\!=\! F_j\forall t\geq
      n\right\},\\\label{eq:Ln} 
    \mbf{C}^j(Z)&=\sup_{n\in\mb{N}}\mbf{C}_n^j(Z).
  \end{align*}
\end{define}
For constants $c_1>0$ and $c_2>0$, consider the optimization problem
\begin{equation}\label{eq:opt_prob}
    \begin{cases}
      \mbf{C}^1(Z)\rightarrow\min\\
      \mbf{C}^0(Z)\rightarrow\min\\
      B^0(Z)\geq c_0\\
      B^1(Z)\geq c_1
    \end{cases}
  \end{equation}
  where $Z$ is a signal process and for $j\in\{0,1\}$ with
  $P\in\mc{P}$ such that $\mc{L}(X_t)=F_j$ for all $t\in\mb{N}$
\begin{equation*}
  B^j(Z)=\inf_{n\in\mb{N}}B_n^j(Z) \hspace{0.25cm}\text{where}\hspace{0.25cm}  B_n^j(Z)=\esssup E \left\{\tau_n^j(Z)-n+1 \mid \mc{F}_{n-1}\right\}.
\end{equation*}
The term $\mbf{C}^1(Z)$ $(\mbf{C}^0(Z))$ is the longest average delay
of signalling out-of-control (in-control), guaranteed regardless of
the distribution of the $X$'s prior to the permanent change
out-of-control (in-control) and prior signals made. Unlike
$\mbf{D}^j(Z)$ which considers the first stopping time $\tau_1^j(Z)$,
we consider the reset stopping time $\tau_n^j(Z)$. Hence it is
possible to repeatedly signal in-control or out-of-control. Each
$B_n^j(Z)$, $n\in\mb{N}$, is the in-control ($j=1$) or out-of-control
($j=0$) average run length after the change point $n$.
\begin{thm}\label{thm:our_opt}
  There exists constants $ k_U>0$
  and $ k_L>0$ such that for any $h\geq\max( k_U, k_L)$ a solution
  of \eqref{eq:opt_prob} is the signal process $Z^*$ defined, for
  $t\in\mb{N}$, by
\begin{equation*}
    Z_t^*=
  \begin{cases}
    0,&R_t^U\leq h- k_U\\
    1,&R_t^L\geq  k_L\\
    \emptyset,&R_t^L<  k_L,\text{ }R_t^U>h- k_U
  \end{cases}.
\end{equation*}
\end{thm}
\noindent The proof of Theorem \ref{thm:our_opt} is in Appendix 1. 

Theorem \ref{thm:our_opt} uses the charts $R_t^L$ and $R_t^U$, which
both follow the same evolutionary equation $f_t$. However, $R_t^L$
starts at value $0$ whereas $R_t^U$ starts at the upper boundary
$h$. Starting at these two values means that we consider both the
cases where the observation start (at time $0$) in the in-control
state and the out-of-control state simultaneously. Thus, the prejudice
of assuming the observations are initially in the in-control state or
the out-of-control state is removed. Further, Theorem
\ref{thm:our_opt} introduces the upper boundary $h$ that prevents the
CUSUM chart rising too high. \cite{gandy11CUSUMFDR} explain how using
an upper boundary in non-restarting CUSUM charts helps detect periods
of out-of-control and in-control activity. Although the upper boundary
$h$ has been introduced in Theorem \ref{thm:our_opt}, its value is not
specified. This practical issue is discussed in detail in Section
\ref{sec:prac} where we explore how varying the upper boundary affects
the signal process.

\section{Upper Boundary}\label{sec:prac}
In this section we investigate how varying the upper boundary $h$
affects the number of false and correct signals made by the charts. We
begin by classifying different scenarios that can occur by changing
$h$. An illustrative example demonstrates how these scenarios leads to
different signals for the same data.

\subsection{Signal Gap and Overlay}\label{sec:gap/overlay}
Setting $h= k_U+ k_L$ in Theorem \ref{thm:our_opt} means that $R_t^L$
and $R_t^U$ signal whenever $ k_L$ is crossed. We shall refer to this
as the single threshold case. However, this need not be the case as
Theorem \ref{thm:our_opt} just stipulates that $h\geq\max( k_U,
k_L)$. When $h > k_U+ k_L$ then it is possible to simultaneously
signal in-control and out-of-control. We shall refer to this case as a
signal overlay. When $h< k_U+ k_L$, then no signal can occur when
either chart is between $h- k_U$ and $ k_L$. We shall refer to this as
a signal gap. Thus both a gap and overlay are intervals where the
charts cannot signal in-control or out-of-control with confidence. If
we consider a simultaneous signal as no signal due to ambiguity, the
cases of signal gap and overlay are the same. Thus, henceforth we
shall only consider $h \leq k_U+ k_L$.

\begin{figure}[tbp]
  \centering 
  \subfloat[$h=6$ Signal
  Gap]{\label{fig:h6}\includegraphics[width=\linewidth]{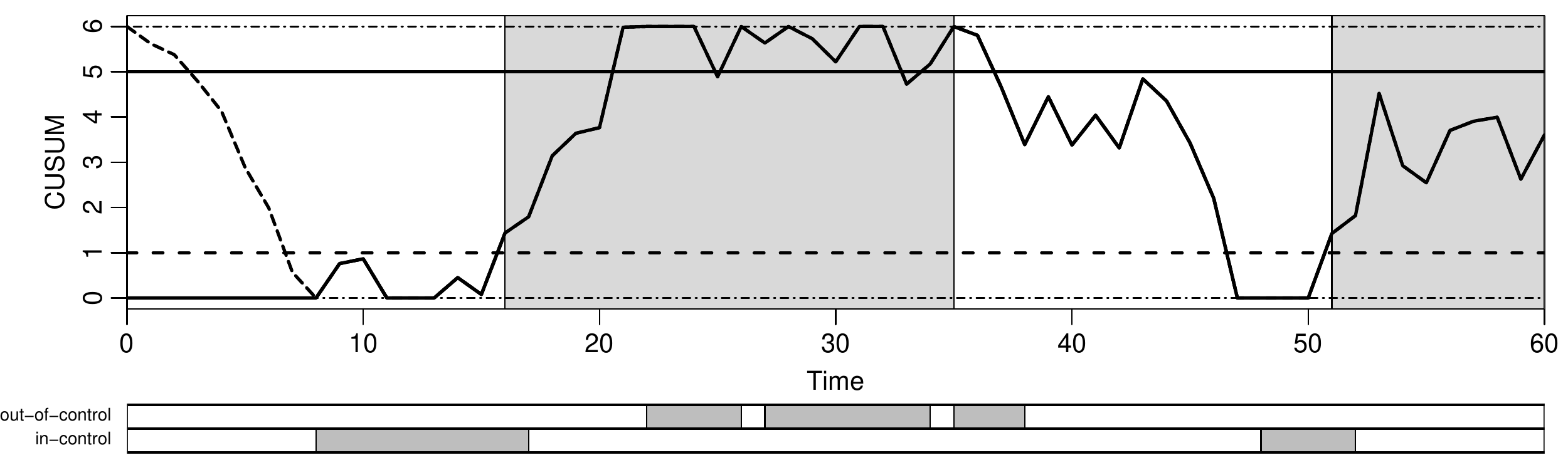}}

  \subfloat[$h=8$ Signal
  Gap]{\label{fig:h8}\includegraphics[width=\linewidth]{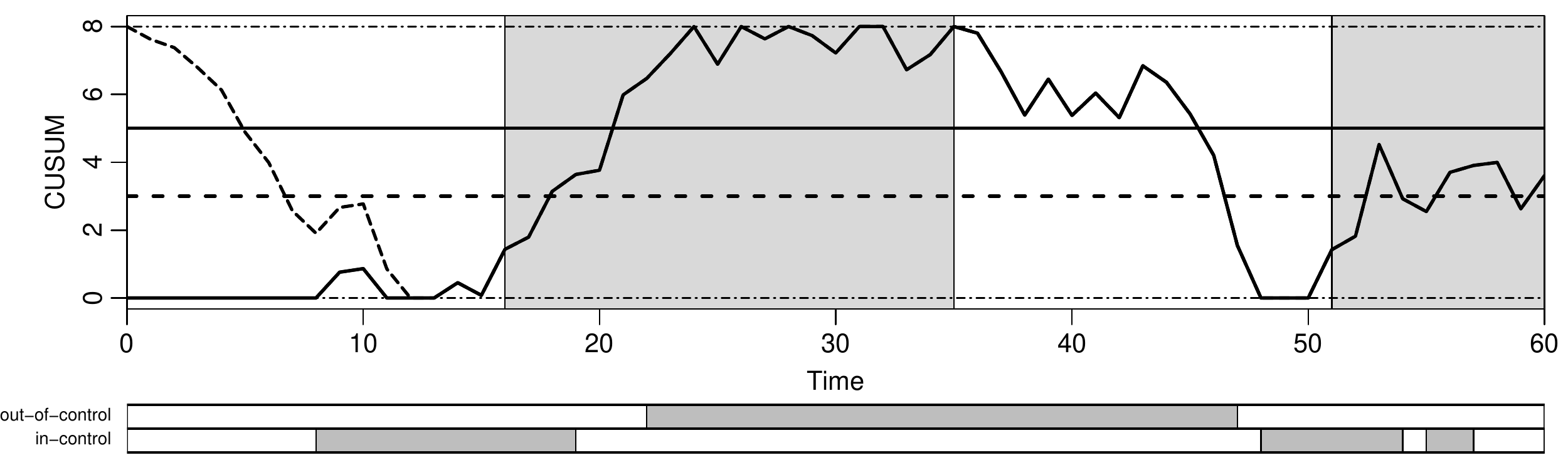}}

  \subfloat[$h=10$ Single
  Threshold]{\label{fig:h10}\includegraphics[width=\linewidth]{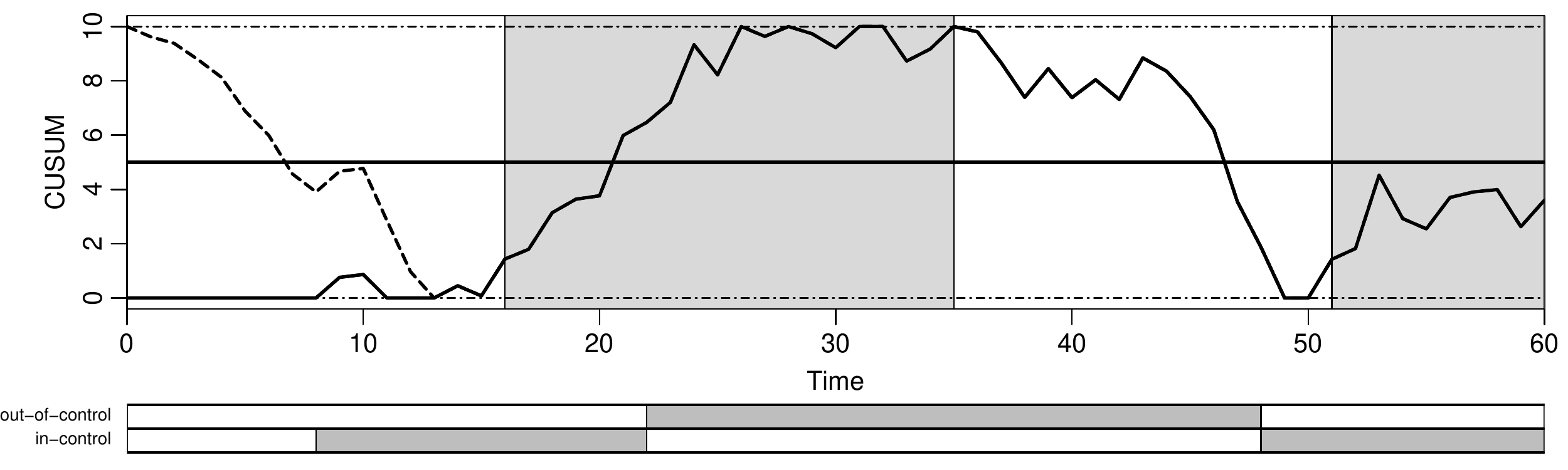}}

  \caption{Upper figures contain one lower (solid) and upper (dashed)
    chart with lower (solid) and upper (dashed) threshold. Grey area
    represents the out-of-control period. Underneath is the signal
    indicator with grey areas representing signals.}
  \label{fig:signals}
\end{figure}

To see the disparity between different values of the upper boundary
$h$ consider the following example. Let $X_t\sim N(-1/2,1)$ when in
the in-control state and $X_t\sim N(1/2,1)$ when in the out-of-control
state at time $t$. In this example we set the thresholds at $ k_U=
k_L=5$. In practice, however, it is typical to set these thresholds by
pre-specifying the average run length. This can be achieved by running
simulations of the CUSUM charts and conducting a numerical
search. Alternatively, one could approximate the distribution of the
CUSUM chart by a discrete Markov chain and obtain the average run
length using the transition matrix \citep{brook}. Using the
\cite{brook} approach with 100 states, the average in-control and
out-of-control run length with $ k_U= k_L=5$ is approximately $930$
time units. In this illustrative example we choose the out-of-control
periods as time $16$ to $35$ and time $51$ to $60$. In Figure
\ref{fig:signals} we plot the two CUSUM charts $R_t^L$ and $R_t^U$
based on the same random realization of $X_1,X_2,\dots$ for $h=6,8$
and $10$.

Figure \ref{fig:signals} illustrates some important points. First
choosing a different value of $h$ leads to a different signalling
process for the same data. This is clearly presented by the signal
indicators under Figure \ref{fig:h6}, \ref{fig:h8} and
\ref{fig:h10}. Second, once the two CUSUM charts have coupled into a
single chart in the single threshold case ($h= k_U+ k_L$), definitive
signals are given. More precisely, in Figure \ref{fig:h10}, after time
$13$, the chart signals either in-control or out-of-control. This is
not the case where a signal gap occurs. Comparing Figure \ref{fig:h6}
and \ref{fig:h8}, we see that as the signal gap increases, fewer
incorrect signals are made at the expense of fewer overall
signals. This point is explored further in Section
\ref{sec:choice}. Lastly, in this example, the two CUSUM charts
eventually coalesce or couple into a single chart. The time until
coupling depends, not only on $F_0$ and $F_1$, but also on the upper
boundary $h$. Coupling occurs at value $0$ or $h$ by construction of
the CUSUM charts. Moreover, if we were the run CUSUM charts of the
form \eqref{eq:our_CUSUM} starting from every $R_0\in[0,h]$ all charts
will couple either at $0$ or $h$. This follows from the fact that
$R_t^L$ and $R_t^U$ represent the ``extreme'' CUSUM charts. Thus at
the time of coupling we signal confidently in-control, if at $0$, or
out-of-control, if at $h$, since all possible charts of the form
\eqref{eq:our_CUSUM} agree.

\subsection{Practical Implications}\label{sec:choice}
Although we have discussed the scenarios that arise with different
choices of the upper boundary, we have yet to explore the practical
implications. As illustrated in Section \ref{sec:gap/overlay}, the
choice of $h$ affects the signals made by the charts. We now explore
how the choice of $h$ affects the number of correct and false signals
in a small simulation.

\begin{figure}[tbp]
  \centering \subfloat[Average Number of False Signals Pointwise in
  Time]{\label{fig:fsignals}\includegraphics[width=\linewidth]{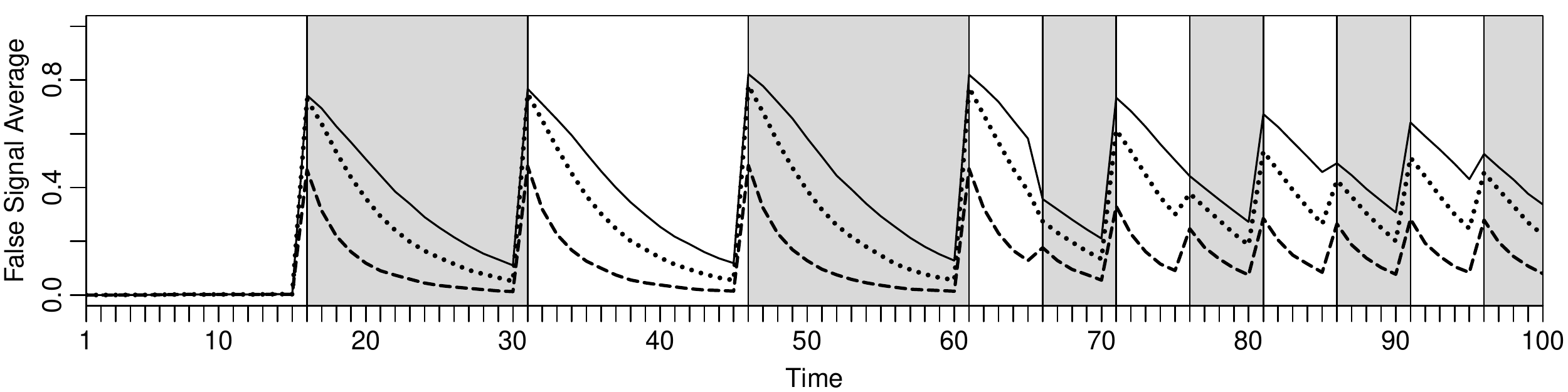}}
  
  \subfloat[Average Number of Correct Signals Pointwise in
  Time]{\label{fig:csignals}\includegraphics[width=\linewidth]{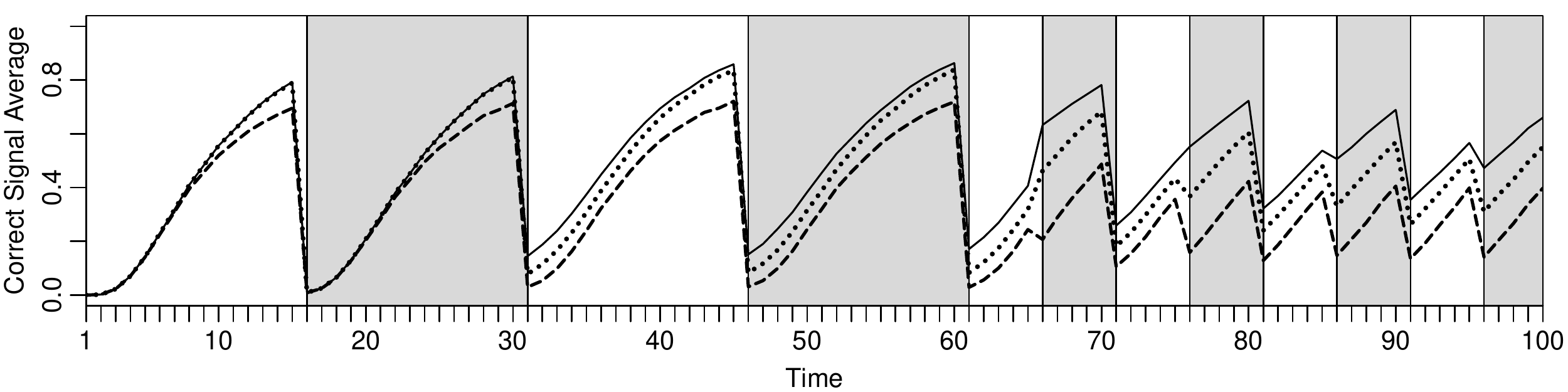}}
  \caption{Average number of false and correct signals for $h=10$
    (solid): same threshold case, $h=8$ (dotted) and $h=6$ (dashed):
    both signal gap cases.}
  \label{fig:fcsignals}
\end{figure}

We use the same in-control and out-of-control distributions and
thresholds used Section \ref{sec:gap/overlay}. The true in-control and
out-of-control periods are represented by the grey areas in Figure
\ref{fig:fcsignals}.
For a single iteration, we use the CUSUM charts $R^L_t$ and $R^U_t$
with $h=6,8,10$ all with the same seed. We repeated this 10,000
times. Figure \ref{fig:fsignals} and \ref{fig:csignals} are plots the
average number of false signals and correct signals, respectively,
pointwise in time.

In Figure \ref{fig:fsignals}, when switching between states, there is
a sudden peak in the number of false signals which subsequently
declines. The single threshold case where $h=10$, leads to the an
overall higher false and correct average than the other values of
$h$. From Figure \ref{fig:fcsignals} the notion of a signal gap can be
interpreted as follows:\ Whilst a chart is in a signal gap, signalling
is deferred until a more ``definitive'' signal can be made i.e.\ when
the chart exits the gap. This results in fewer false signals being
made, with the trade-off that fewer signals overall are made. In
practice, this could represent the users attitude to ``err on the side
of caution'' and patiently waiting to signal with more confidence.

\section{Discussion}\label{sec:discussion}
We have shown that using non-restarting CUSUM charts of the form
\eqref{eq:our_CUSUM} are optimal in the sense of
\eqref{eq:opt_prob}. We specifically use the two charts, $R^L_t$
starting at $0$ and $R_t^U$ starting at the upper boundary. This
choice results in coupling of the CUSUM charts (see Figure
\ref{fig:signals}) where after the charts coalesce, they remain
exactly the same. It is clear that after coupling, only one CUSUM
chart needs to be considered.Coupling would not occur when using two
CUSUM charts that are restarted - this feature is specifically for
non-restarting CUSUM charts with an upper boundary.

A natural question to ask about coupling is:\ how long do the two
non-restarting CUSUM charts take to couple? Denote the coupling time
as $T=\min\{t\geq 1 : R_t^L=R_t^U \}$. Denote
$\nu_\uparrow=\min\{t\geq 1 : R_t^L=h \}$ and
$\nu_\downarrow=\min\{t\geq 1 : R_t^U=0 \}$ as the first time when the
lower CUSUM chart reaches the upper boundary and the upper CUSUM chart
reaches $0$ respectively. As $T=\min(\nu_\uparrow,\nu_\downarrow )$ it
follows that $E(T) \leq \min(E(\nu_\uparrow),E(\nu_\downarrow))$ for
all $P\in\mc{P}$, since all CUSUM charts of the form
\eqref{eq:our_CUSUM} with any starting value, couple either at value
$0$ or the upper boundary. For $P\in\mc{P}$ such that
$\mc{L}(X_t)=F_1$ for all $t\in\mb{N}$ we would expect
\begin{equation*}
E(T)\leq E(\nu_\uparrow)\ll E(\nu_\downarrow)
\end{equation*}
and similarly for $P\in\mc{P}$ such that $\mc{L}(X_t)=F_0$ all
$t\in\mb{N}$,
\begin{equation*}
  E(T)\leq E(\nu_\downarrow)\ll E(\nu_\uparrow).
\end{equation*}
Proving such coupling results could be a topic for
further research.

\appendix
\section*{Appendix 1}
\section*{Proof of Theorem \ref{thm:our_opt}}\label{ap:proof1}
We proceed to show the following:\ for any signal process $Z$ and
$j\in\{0,1\}$
\begin{equation*}
  \mbf{C}^j(Z)\geq \mbf{D}^j(Z)\geq \mbf{D}^j(Z^j)=\mbf{C}^j(Z^*).
\end{equation*}
We start by showing $\mbf{C}^1(Z)\geq \mbf{D}^1(Z)$ for any signal
process $Z$. As a first step we show that
\begin{align}\label{eq:s1.5}
   \mbf{C}^1(&Z)\geq\\\nonumber
  \sup_{n\in\mb{N}}\max& \left\{\esssup E \left(\left[\tau_1^1(Z)-n+1\right]^{+}\mid \mc{F}_{n-1}  \right):P\!\in\!\mc{P},\mc{L}(X_t)\!=\! F_1\forall t\geq n \right\}.
\end{align}
Consider the stopping time in $\mbf{C}_n^1(Z)$, namely $\tau_n^1(Z)-n+1$
and $\left[\tau_1^1(Z)-n+1 \right]^{+}$. For all $j\geq 1$ and fix
$n\in\mb{N}$ we have
\begin{equation*}
  \{\tau_n^1(Z)-n+1 = j  \} =   \{Z_m \neq 1 \text{ for } n \leq m<(j+n-1), Z_{j+n-1}=1  \}
\end{equation*}
and
\begin{equation*}
  \{[\tau_1^1(Z)-n+1]^{+} = j  \} =    \{Z_m \neq 1 \text{ for }1\leq m < (j+n-1), Z_{j+n-1}=1  \}.
\end{equation*}
Thus $\{[\tau_1^1(Z)-n+1]^{+} = j \}\subseteq \{\tau_n^1(Z)-n+1 = j
\}$ and so have a pointwise order of the stopping times. This implies
\eqref{eq:s1.5}.

Next we show that
\begin{align}\nonumber
 \sup_{n\in\mb{N}}\max&\!\left\{\esssup E\!\left(\left[\tau_1^1(Z)-n+1\right]^{+}\!\mid \mc{F}_{n-1}  \right): P\!\in\!\mc{P},\mc{L}(X_t)\!=\! F_1\forall t\geq n \right\}\\\label{eq:s1.75}
 &\geq \mbf{D}^1(Z).
\end{align}
For a fixed $n\in\mb{N}$
\begin{align}\label{eq:inter}
  \max&\left\{\esssup E\!\left(\left[\tau_1^1(Z)-n+1\right]^{+}\!\mid
      \mc{F}_{n-1} \right): P\!\in\!\mc{P},\mc{L}(X_t)\!=\! F_1\forall
    t\geq n \right\}\\\nonumber
  &\geq \esssup E\!\left(\left[\tau_1^1(Z)-n+1\right]^{+}\!\mid \mc{F}_{n-1}  \right)=\mbf{D}_n^1(Z) 
\end{align}
where, on the right hand-side of the inequality, is under $P\in\mc{P}$
such that $\mc{L}(X_t)=F_0$ for all $t<n$ and $\mc{L}(X_t)=F_1$ for
all $t\geq n$. Taking the supremum of \eqref{eq:inter} over
$n\in\mb{N}$ shows \eqref{eq:s1.75}. This completes showing
$\mbf{C}^1(Z)\geq \mbf{D}^1(Z)$. A similar argument can be used to
show $\mbf{C}^0(Z)\geq \mbf{D}^0(Z)$.

Since $B^j(Z)\geq c_j$ it follows that $B_1^j(Z)\geq c_j$ for
$j\in\{0,1\}$. As $B_1^j(Z)$ is the expectation in \eqref{eq:mous} it
follows that \citep[][Proposition 2]{Ritov1990} $\mbf{D}^j(Z)\geq
\mbf{D}^j(Z^j)$ for any signal process $Z$.

A consequence of \citet[Lemma 1]{moustakides} is that
$\mbf{D}^1(Z^1)=\mbf{D}_1^1(Z^1)$. Thus, by definition of
$\mbf{D}_1^1(Z^1)$,
\begin{equation}\label{eq:a}
  \mbf{D}^1(Z^1)=E\left(\inf\{t\geq 1 : S_t^L\geq k_L\}\mid P\in\mc{P}:\mc{L}(X_t)=F_1\forall t\in\mb{N} \right).
\end{equation}
Replacing $\inf\{t\geq 1 : S_t^L\geq k_L \}$ in \eqref{eq:a} with
$\inf\{t\geq 1 : R_t^L\geq k_L \}$ as they are the same, gives
\begin{equation*}
  \mbf{D}^1(Z^1)=\mbf{C}^1_1(Z^*).
\end{equation*}
Thus, to show $\mbf{D}^1(Z^1)=\mbf{C}^1(Z^*)$, it suffices to show
$\mbf{C}^1(Z^*)=\mbf{C}_n^1(Z^*)$ for all $n\in\mb{N}$.

To show this we follow an argument similar to that used in
\citet[Lemma 1]{moustakides}. For any $m>n\geq 1$ and for fixed
$\{X_{n+1},\dots,X_m\}$, the quantity $R_m^L$ is a non-decreasing
function of $R_n$. This implies that the stopping time
$\tau_n^1(Z^*)=\inf\{t\geq n : R_t^L\geq k_L\}$ is non-increasing with
$R_{n-1}$. Thus the essential supremum in $\mbf{C}_n^1(Z^*)$ is
achieved for $R_{n-1}=0$. Moreover, the essential supremum in
$\mbf{C}_n^1(Z^*)$ remains unchanged when taking the maximum over
$P\in\mc{P}$ such that $\mc{L}(X_t)=F_1$ for $t\geq n$. Hence from
stationarity all $\mbf{C}_n^1(Z^*)$ are equal. This completes showing
$\mbf{D}^1(Z^1)=\mbf{C}^1(Z^*)$.

We now show that $\mbf{D}^0(Z^0)=\mbf{D}^0(Z^*)$. Similarly, we have
$\mbf{D}^0(Z^0)=\mbf{D}_1^0(Z^0)$ thus by definition of
$\mbf{D}_1^0(Z^0)$
\begin{equation}\label{eq:b}
  \mbf{D}^0(Z^0)=E(\inf\{t\geq 1 :S_t^U \geq k_U\}\mid P\in\mc{P}:\mc{L}(X_t)=F_0\forall t\in\mb{N}).
\end{equation}
Replacing $\inf\{t\geq1 : S_t^U\geq k_U \}$ in \eqref{eq:b} with
$\inf\{t\geq1 : R_t^U\leq h-k_U \}$ as they are the same, gives
$\mbf{D}^0(Z^0)=\mbf{C}_1^0(Z^*)$. Showing that
$\mbf{C}^0(Z^*)=\mbf{C}^0_n(Z^*)$ for all $n\in\mb{N}$ follows from
the same argument given above. Thus $\mbf{D}^0(Z^0)=\mbf{C}^0(Z^{*})$.

\bibliography{bibil}
\end{document}